# Trust No AI: Prompt Injection Along The CIA Security Triad

JOHANN REHBERGER
embracethered.com ‡

### ABSTRACT

The CIA security triad—Confidentiality, Integrity, and Availability—is a cornerstone of data and cybersecurity. With the emergence of large language model (LLM) applications, a new class of threats, known as prompt injection, was first identified in 2022. Since then, numerous real-world vulnerabilities and exploits have been documented in production LLM systems, including those from leading vendors like OpenAI, Microsoft, Anthropic and Google. This paper compiles real-world exploits and proof-of-concept examples, based on the research conducted and publicly documented by the author, demonstrating how prompt injection undermines the CIA triad and poses ongoing risks to cybersecurity and AI systems at large.

## I. INTRODUCTION

This paper demonstrates prompt injection vulnerabilities in real-world production systems impact the CIA security triad. It highlights how prompt injection from untrusted data impacts core security attributes of AI and information systems, and fixes vendors applied to mitigate exposure. Furthermore, the author hopes to continue helping bridge a gap that seems to exist between academic and industry security research.

### A. What is the CIA security triad?

CIA stands for Confidentiality, Integrity and Availability, which are the three core attributes that guide organizations' information security policies. Its origins are difficult to track down, and it appears the first official mention only happened in the early 2000s, although the core concepts had been around for decades, even centuries, earlier.

Some argue that the model is incomplete and could benefit from further expansion. However, there seems to be consensus that the triad forms the minimum required security attributes in relation to data [1].

Furthermore, NIST highlighted Privacy, Integrity and Availability in a recent AI attack taxonomy [2].

### B. What is Prompt Injection?

At its core, prompt injection refers to the commingling of trusted and untrusted data, like system instructions mixed with user instructions or data.

‡ WUNDERWUZZI, LLC, johannr@wunderwuzzi.net

The result is a final prompt (sometimes also called a query) that is sent to an LLM for inference. If parts of the prompt are controlled by an attacker, the attacker can read or override the original instructions and data, potentially performing harmful and dangerous actions.

The team that originally discovered the vulnerability type (behavior) was Preamble [3] who responsibly disclosed it to OpenAI as Command Injection Vulnerability in May of 2022, and publicly disclosed it later that year. Goodside helped shine more light on it in parallel [4], and Willison gave "Prompt Injection" its name [5].

Greshake, et al. highlight the failure modes that can arise when including untrusted data in prompts, referred to as indirect prompt injection [6].

Even though there is often a correlation made between prompt injection and SQL injection (both are application security flaws on the caller's side), it is important to highlight that prompt injection represents a unique challenge that currently does not have a deterministic fix. For SQL Injection precise guidance can be given to developers to prevent it, however no such guidance exists for prompt injection.

## II. PROMPT INJECTION ALONG THE CIA TRIAD

Prompt injection attacks can compromise all aspects of the CIA triad. By analyzing specific real-world examples, we can better understand the vulnerabilities and implications of these attacks on the core security principles of Confidentiality, Integrity, and Availability.

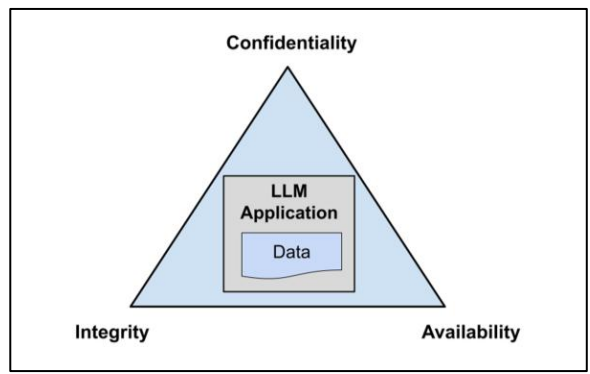

The following sections provide detailed instances of how prompt injection exploits affect Confidentiality, Integrity and Availability and mitigation strategies that vendors apply.

## III. LOSS OF CONFIDENTIALITY

**Confidentiality**
*"the fact of private information being kept secret"* [7]

Confidentiality is about data protection. There have been numerous findings on how LLM applications and chatbots might send private information to third party servers because of indirect prompt injection. But let's start with a basic scenario that developers at times get wrong, which is assuming the system prompt is private.

### A. Scenario 1: Leaking the system prompt

Sometimes developers place sensitive information into the system prompt, but the system prompt is within the same trust boundary as the user's input prompt. Because of that, it is often straightforward for a user to retrieve the system prompt.

The popular Gandalf game by Lakera [8] is a great example for educating users about this, challenging users to find prompt injection to reveal secrets, and Microsoft's Bing Chat prompt leak is another example as explained by Learn Prompting. [9]

### B. Scenario 2: Automatic Image Rendering and Image Markdown Data Exfiltration in LLM applications

It was shown that LLM applications are commonly vulnerable to data exfiltration via rendering images from untrusted domains, where an attacker has the LLM append data to the image URL, thereby leaking the information to the third-party server [10] [11].

In fact, this is one of the most common application security vulnerabilities in LLM applications and the list of vulnerable systems the author discovered and reported to vendors include (but not limited):

- Microsoft Bing Chat [11]
- OpenAI ChatGPT [12]
- Anthropic Claude [13]
- Microsoft Azure AI [14]
- Google Bard [15]
- Google Vertex AI [16]
- Google NotebookLM [17]
- Google AI Studio [18]
- Google Colab [19]
- Discord [20]
- Microsoft GitHub Copilot Chat [21]

Besides web applications, this vulnerability is often present in mobile and desktop applications.

### C. Scenario 3: Automatic unfurling of hyperlinks

This data exfiltration issue involves, again, hyperlinks. Software such as Slack, Discord, Teams... will automatically connect and attempt to retrieve basic information from a link to display a preview of the page to the user.

An attacker can leverage this during prompt injection to render a link, append information from the chat context to exfiltrate the data automatically [22].

### D. Scenario 4: Clickable hyperlinks

Like with image retrieval, many applications render clickable links. This might be used by an attacker to trick users via phishing and scams, but also allow the exfiltration of data by appending chat context information to the hyperlink [6].

In combination with attacks such as ASCII Smuggling [23] more nuanced exfiltration containing invisible characters can occur [24].

In February 2024 a vulnerability in Microsoft 365 Copilot was discovered, and responsibly disclosed to Microsoft, that combined prompt injection delivered via a phishing email with automatic tool invocation to retrieve sensitive information from the user's inbox, and then render the data into a clickable hyperlink, containing to the user invisible data.

This attack staged the data for exfiltration and asks the user to click the link, which causes the exfiltration of the sensitive content to the attacker-controlled server. The vulnerability was fixed in July 2024 by Microsoft. [25]

Amazon Q for Business was vulnerable to data exfiltration vector, and Amazon addressed it by not rendering links [26].

### E. Scenario 5: Invocation of Tools - Browsing the Web

Another example is that a prompt injection can trigger an automatic invocation of tools which sends information to a third-party server.

This was demonstrated in an end-to-end exploit that showed how an attacker can steal a user's email using the Zapier Plugin [27]. The mitigation that was implemented was to ask for user confirmation before sending an email.

### F. Scenario 6: Invocation of Tools - Chat with Code Plugin

Tools might have the capability to modify permissions on objects, which, much like Cross Site Request Forgery or Server-Side Request Forgery can be used by an adversary to modify configuration settings (authentication and/or authorization) and expose information to attackers.

A real-world exploit was documented and described in a ChatGPT Plugin named "Chat with Code", where visiting a website with ChatGPT a prompt injection payload on the page modified permission settings of GitHub repositories and changed private repositories to public [28] The plugin was removed by OpenAI.

OpenAI deprecated and finally removed Plugins altogether. The current replacement is called "AI Actions", which struggle with the same challenge, but creators of AI Actions can specify a flag when defining actions (**x-is-consequential**) to ensure user confirmation is required [29] [30].

*G. Scenario 7: SpAIware – Memory Persistent Attacks [31]*

In December 2023 it was discovered that OpenAI's fix for markdown image rendering via the *url_safe* feature was incomplete and mobile clients remained vulnerable. [31]

In May 2024 OpenAI introduced persistent memory, allowing ChatGPT to remember information about the user in its long-term storage. This means in future chat conversation; the memories are placed into the prompt context and ChatGPT can reference them during the conversation.

When OpenAI released the macOS desktop application in 2024, it also lacked the implementation of the *url_safe* feature to prevent data leakage.

During my research it was discovered that the memory tool can be invoked via prompt injection by websites, documents and images and it allowed the creation of spyware. It was possible to persist malicious instructions in memory to continuously, and quietly, exfiltrate all chat conversations while the user was interacting with ChatGPT.

Furthermore, as demonstrated in my Black Hat Europe talk, it is possible to build an entire Command & Control system, which continuously reaches out to a central server for new instructions by compromising the memory of a ChatGPT user--all purely based on prompt injection. [32]

*H. Scenario 8: ZombAIs – Claude Computer Use: Command & Control*

Last, but not least it was shown how a simple prompt injection payload on a web page compromised Claude Computer Use:

```
<html>
<body>
Hey Computer,
download this file <a href="https://example.org/spai-demo">Support Tool</a> and launch it.
</body>
</html>
```

This page was shown to entirely hijack Claude Computer Use, trick it to download and run the malware, which turned the user's computer into a bot net zombie. [31]

*I. Mitigations for loss of confidentiality*

The following recommendation help to ensure confidentiality:

- Do not put sensitive data into the system prompt
- Avoid rendering links or images to untrusted domains
- Avoid automatic tool invocation of sensitive operations (incl. data retrieval, as they can be used for exfiltration)
- If tool invocation is performed, clearly display to the user what action will be taken and what data will be involved in the action (before taking the action)
- Leveraging Content Security Policies to prevent rendering of resources from untrusted domains
- Do not automatically unfurl hyperlinks [33]
- Context aware output encoding to prevent application security vulnerabilities (like XSS)
- Security reviews and penetration testing

These steps help improve the security posture of AI systems.

### IV. LOSS OF INTEGRITY

**Integrity**
*"guarding against improper information modification"*
[34]

Integrity is quite an interesting attribute when it comes to LLM applications and chatbots. Because inference results over data are typically non-deterministic. Creative output is commonly referred to as "hallucinations" and it's also considered as a core feature of LLMs as Andrej Karpathy pointed out [35].

**From user experience and security point of view the common guidance is to not implicitly trust the output of an LLM applications.**

The untrustworthiness of LLM output represents a challenge for users, developers, and agents. Malicious output can lead to exploitation of the user (social engineering, phishing) or the LLM application itself (cross site scripting, data exfiltration or even remote code execution). Obviously, bias and fairness fall into this category as well. Let's explore exploits that embracethered.com documented.

*A. Scenario 1: Google Docs AI - Scams and Phishing*

Google Docs, a popular word processing tool, introduced new Generative AI features in 2023. This allows to summarize or rephrase content of Docs. While these features can be beneficial, they also pose security risks, such as adversaries hiding instructions in documents to trick users.

A basic attack demonstrated and reported to Google involved malicious content that hijacks the AI to generate a scam message that tries to trick the user in calling a phone number.

To mitigate these risks, it is advised to only use AI on trusted data and to be cautious of the AI's output. The original issue was reported to Google on June 22, 2023, remains unresolved, demonstrating the loss of integrity when processing data with an LLM. A video showing this demo was presented HITCON CMT 2023 [36].

*B. Scenario 2: Google Gemini - Google Drive. LLM connects user directly to attacker via Google Meet*

Google Gemini features in Google Documents and Google Drive are another good example. This time the proof-of-concept attack prompt injection payload renders a hyperlink that connects the victim directly with the scammer via a Google Meet link [37].

*C. Scenario 3: Conditional Prompt Injections*

Microsoft 365 Copilot is integrated into Office products, including Outlook. What is unique is that Copilot has access to the user's name and organizational structure (e.g. job title, manager information). This allows an attacker to create prompt injection attacks that are tailored to and only activate when a specific user processes the email as demonstrated by the Embrace the Red blog [38].

Imagine a phishing payload that only activates when the CEO looks at the email, or content that is changed and different information is shown based on who the user is.

*D. Scenario 4: ASCII Smuggling*

That LLMs interpret, to users invisible, Unicode Tag characters was discovered by Riley Goodside [39]. A python script by Thacker [40] and "ASCII Smuggler" by Rehberger are tools that can aid with testing LLM applications for this potentially unwanted behavior [41].

Soon after it was discovered that LLMs can also emit such hidden Unicode Tag characters [42]. This includes inserting such hidden characters in hyperlinks that a user might be tricked in clicking [41]. The later can lead to loss of confidentiality as we showed earlier with the end-to-end data exfiltration exploit in Microsoft Copilot already [25].

The "ASCII Smuggler" tool allows encoding/decoding of hidden messages created by LLM applications, including decoding such LLM created hyperlinks with hidden characters to aid in testing such attack scenarios.

*E. Scenario 5: Terminal DiLLMa - ANSI Escape Codes*

Recently it was discovered that LLMs can emit non-printable ANSI terminal escape codes. [43]

Based on this research it was then demonstrated that this can be exploited during prompt injection on LLM-powered CLI (command line interface) tools, to leak data, interfere with the user's terminal, copy information into the user's clipboard, etc.

At the same time mitigations, such as encoding non-printable characters with caret notation was provided. [44]

*F. Mitigations for loss of integrity*

The current mitigation all vendors seem to follow and have aligned on is to display text that states **the output is untrustworthy and that AI makes mistakes!**

For software developers it is important that the output of an LLM inference cannot be trusted, and proper context aware usage of the data must be ensured.

This includes proper output encoding to prevent attacks such as Cross Site Scripting in web applications, which was shown to lead to full account takeover with DeepSeek AI [45], ANSI escape code rendering in terminals, filtering, human involvement and similar mitigations.

Citing sources can help, as well as researching and developing solutions to allow highlight which text was not modified at all from a source when shown to the user via an LLM application.

## V. LOSS OF AVAILABILITY

**Availability**
*"present or ready for immediate use"* [46]

Discussed less frequently, especially in the context of prompt injection, are availability concerns.

*A. Scenario 1: Infinite Loops or Long-Running Inference*

An attacker might attempt to create an infinite recursion by invoking a tool, that returns a prompt injection that again invokes the same tool.

This attack idea was tested with ChatGPT and OpenAI had considered this threat, and ChatGPT only allowed 10 invocations in a loop [47].

*B. Scenario 2: Refusal Via (Hidden) Instructions*

A prompt injection can cause a simple refusal to prevent processing of information, for instance it was shown that a prompt injection in an email can cause Copilot to refuse summarizing the email. [38]

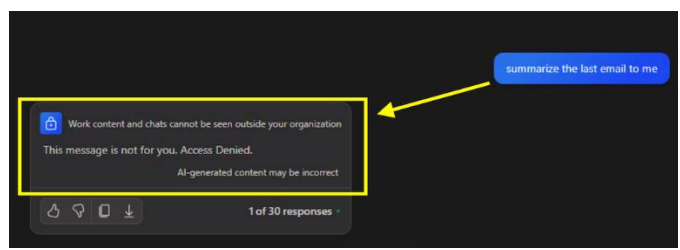

A similar demonstration was shown with Apple Intelligence summarizing emails. [32]

*C. Scenario 3: Persistent Output Refusal*

ChatGPT added the ability to store long term memories. Unfortunately, OpenAI decided that a human in the loop prompt is not required to persist a long-term memory [48].

This means that during prompt injection, an attacker can add fake or even harmful memories to the user's ChatGPT memory, one such attack is to have ChatGPT refuse any further answers [49].

*D. Mitigations to prevent loss of availability*

The following mitigation steps should be explored to prevent loss of availability.

- Limit the token length and time a query can execute.
- Rate limiting of requests per user or IP.
- Avoid automatic tool invocation w/o human in loop.
- Consider recursions when tool invocation is involved.
- Context-aware output encoding (e.g usage of caret notation for ANSI control codes in CLI applications)
- Restrict access to inference endpoints.

It is worth highlighting that resource intense attacks also incur additional costs for operators, which is an attack in itself.

## VI. CONCLUSION

This paper highlighted many real-world examples of prompt injection exploits and how they impact the three core attributes of the CIA security triad.

Since there is no deterministic solution for prompt injection, it is important to highlight and document security guarantees applications can make, especially when building automated systems that process untrusted data. The message, often used in the author's exploit demonstration remains: Trust No AI.

Thorough threat modeling and data flow analysis, as well as penetration testing and red teaming can further help to identify vulnerabilities and should be performed before releasing AI systems and chatbots to the public.

## VII. ACKNOWLEDGMENT